\documentclass[reprint, final, aps, prl]{revtex4-1}

\usepackage{graphics}
\usepackage{graphicx}
\usepackage{epsfig}
\usepackage{amsthm}
\usepackage{amsmath}
\usepackage{amssymb}
\usepackage{lineno}
\usepackage{fleqn}
\usepackage{txfonts}
\usepackage{url}
\usepackage{bm}

\usepackage{epstopdf}
\usepackage{float}

\begin{document}
\title{Robust synchronization of an arbitrary number of spin-torque driven vortex nanooscillators}

\author{Sergey Erokhin}
\author{Dmitry Berkov}
\email[]{db@innovent-jena.de}
\affiliation{Innovent Technology Development, Pr\"ussingstra{\ss}e 27B, D-07745, Jena, Germany}

\date{\today}

\begin{abstract}
Non-linear magnetization dynamics in ferromagnetic nanoelements excited by the spin-polarized dc-current is one of the most intensively studied phenomena in solid state magnetism. Despite immense efforts, synchronization of oscillations induced in several such nanoelements (spin-torque driven nanooscillators, or STNO) still represents a major challenge both from the fundamental and technological points of view. In this paper we propose a system where synchronization of any number of STNOs, represented by magnetization vortices inside squared nanoelements, can be easily achieved. Using full-scale micromagnetic simulations we show that synchronization of these STNOs is extremely dynamically stable due to their very large coupling energy provided by the magnetodipolar interaction. Finally, we demonstrate that our concept allows robust synchronization of an arbitrary number of STNOs (arranged either as a 1D chain or as a 2D array), even when current supplying nanocontacts have a broad size distribution.
\end{abstract}

\maketitle

\section{I. INTRODUCTION}

Magnetization excitation and switching in thin ferromagnetic nanostructures induced by a spin-polarized dc-current is a very active research area due to the large importance of these phenomena for our fundamental understanding of electromagnetic properties of condensed matter \cite{hmm_vol_5, Ralph_JMMM_2008}. Particularly, an enormous attention has been paid to spin torque excitation of a steady state gyration of magnetization vortices in these systems. This object is a highly non-trivial example of a nanosized oscillator with strongly nonlinear properties. For this reason the study of its dynamical behavior, leading to the decisive improvement of a general understanding of non-linear effects in magnetic systems, is one of the most important research tasks in non-linear magnetic phenomena. Especially various synchronization regimes in ensembles of these oscillators represent a very challenging fundamental problem, which solution for various cases has been the subject of many recent publications \cite{Slavin_PRB_2006, Tiberkevich_APL_2009, ChenXi_PRB_2009, DongLi_PRB_2011, Belanovsky_PRB_2012}.

This interest is  strongly supported by many promising applications of this effect \cite{Silva_JMMM_2008, Katine_JMMM_2008}. One of the most important among them would be a nanosized microwave generator which frequency could be tuned by changing either the external magnetic field or the dc-current strength. Such generator could be very useful in, e.g., detectors of microwave fields and telecommunication appliances.

Despite large efforts aimed to create such generators, the power and linewidth of a single STNO are still not competitive with their non-magnetic state-of-the-art analogues. Decisive improvement of STNO-based systems is expected from synchronization (phase-locking) of several STNOs, what would result both in a quadratic increase of the output power ($\sim N^2$ for $N$ synchronized oscillators) and substantial narrowing of the generated linewidth. Unfortunately, although intensive  theoretical research led a substantial deepening of our understanding of STNO synchronization process \cite{Slavin_PRB_2005, Slavin_PRB_2006, Rezende_PRL_2007, Tiberkevich_APL_2009, ChenXi_PRB_2009, DongLi_PRB_2011, Belanovsky_PRB_2012}, experimental achievements  remain rather modest: in pioneering reports \cite{Kaka_Nat_2005, Mancoff_Nat_2005} two closely placed point contact STNOs could be synchronized, and afterwards only in one experimental paper \cite{Ruotolo_Nat_2009} the synchronization of four such oscillators was demonstrated.

In most studies synchronization of STNOs with a nearly collinear magnetization configuration within the contact area is considered. Only recently it has been shown that a steady-state vortex gyration \cite{Mistral_PRL_2008, Darques_JPhysD_2011, Petit-Watelot_NatPhys_2012, Jaromirska_PRB2011} is also a possible candidate for realizing a system of synchronized STNOs \cite{Ruotolo_Nat_2009, Belanovsky_PRB_2012}. 

We note here that the dynamics of vortex pairs in nanoelements coupled via the dipolar \cite{Sugimoto_PRL2011, Jung_SciRep_2011} and exchange \cite{Buchanan_NatPhys_2005} interaction, as well as dynamics of one-dimensional \cite{Sukhostavets_PRB2013} and two-dimensional \cite{Vogel_PRL2010, Sukhostavets_PRB2013} vortex arrays has been studied very intensively during last decade both theoretically and experimentally (see also references in \cite{Guslienko_JNanosci_2008, Antos_JPSJ_2008}). In all these publications vortex dynamics excited by a time-dependent external magnetic field has been investigated. In contrast to these numerous studies there exists - up to our knowledge - only a single publication \cite{Belanovsky_PRB_2012} devoted to the synchronization of two vortex STNOs (with vortices inside two adjacent nanodisks) under the influence of a spin-polarized current.

In this paper we present a systematic study of the synchronization of vortex STNOs in various patterned thin film structures. The majority of results presented below is obtained for systems with vortices inside square-shaped nanoelements. We begin with the analysis of a single vortex dynamics within such a nanosquare in an external field directed perpendicular to the nanoelement plane (Sec. II). In Sec. III we study in detail synchronization of vortices in a pair of adjacent nanosquares, which are either (i) fully separated from each other (i.e. the interaction between nanoelements is purely dipolar), or (ii) connected via a 'bridge' made of the same magnetic material (i.e., dipolar and partially exchange coupled vortices) or (iii) fully connected, so that the system under study is actually a system of two vortices inside a rectangular nanoelement. The major results of this Section are that such STNOs can be synchronized in a broad current range and that due to a very high coupling energy between the constituent STNOs, their synchronization is stable with respect to thermal fluctuations. In Sec. IV we present the system concept which allows the robust synchronization of an {\it arbitrary} number of vortex-based STNOs - arranged in a chain - in a wide range of currents, even when point contacts (through which the spin-polarized current is injected) have a broad distribution of their diameters. In Sec. V we expand our concept to a 2D array of vortices in magnetic nanosquares. In both Sec. IV and V we explicitly show that due to the absence of any phase difference between the vortex STNOs we can achieve a perfect $\sim N^2$ scaling of the oscillation power with the number of oscillators $N$. Our conclusions are given in Sec. VI.

\section{II. MAGNETIZATION DYNAMICS OF A SINGLE VORTEX IN A NANOSQUARE}

To provide a basis for understanding of our many-vortices system, we start with the analysis of magnetization dynamics for a single vortex-based STNO. To build a vortex we have used the $500 \times 500$ nm$^2$ square-shaped nanoelement with the thickness $h = 20$ nm, what corresponds to a typical experimental system (see, e.g., \cite{Petit-Watelot_NatPhys_2012}). The current with the out-of-plane spin polarization direction and polarization degree $P = 0.3$ is injected via the central point contact with the diameter $D=100$ nm (unless stated otherwise, see Sec. IV and V). Magnetic parameters typical for Permalloy are chosen: magnetization $M = 860$ G, exchange constant $A = 1.0 \times 10^{-6}$ erg/cm, and Gilbert damping $\lambda = 0.01$. To prevent the reversal of the core polarity for high currents, the constant external field $H_{\rm perp} = 2$ kOe was applied perpendicularly to the nanoelement plane. The system was discretized only in-plane using a $4 \times 4$ nm$^2$ mesh. Micromagnetic simulations were carried out with our MicroMagus software package \cite{MicroMagus}, which dynamical part solves the Landau-Lifshitz-Gilbert equation of motion for magnetic moments including the additional spin-torque term. For all systems studied here magnetization dynamics was simulated during 200 ns. In order to explore only the steady precession regime, all spectra shown below have been calculated from the data collected for $ t > 100$ ns.

The next important issue is how to initiate the vortex gyration, because in presence of the circularly symmetric Oersted field, the state with the minimal energy corresponds to the vortex core position in the nanocontact center. Thus, to induce vortex oscillations, an in-plane field pulse with the amplitude 200 Oe and duration 3 ns was applied. This pulse was strong enough to 'knock out' the vortex into the region outside the contact. In this region the competing influences of the spin torque, Gilbert damping, Oersted field and the stray field from the nanosquare borders led to the vortex motion around the point contact {\it outside} the contact area. After the system 'forgets' the initial in-plane field pulse, this rotation occurs by a nearly circular orbit.

\begin{figure}[]
\includegraphics[width=90mm]{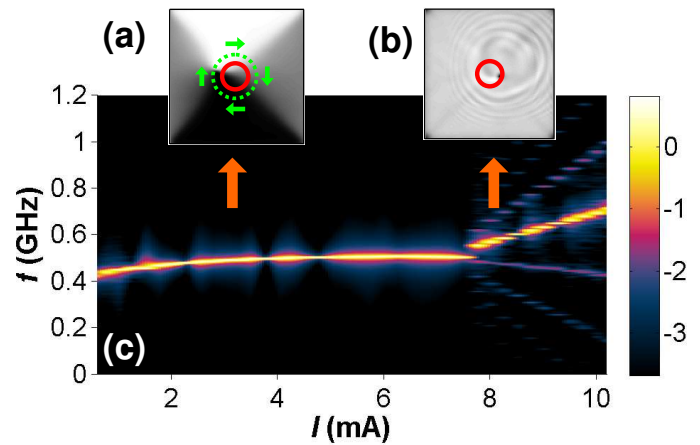}
\caption{(color online) Magnetization dynamics of a single STNO: (a) and (b) - snapshots of magnetization configurations in low- and high-current oscillation regimes (solid circles define the nanocontact area, the dotted line in (a) is the vortex trajectory); (c) spectral amplitude of magnetization oscillations under the contact.
\label{SingleVortex_Power}}
\end{figure}

Simulation results for our single nanooscillator are shown in Fig. \ref{SingleVortex_Power}. Note, that on this and all other figures we display the spectral amplitude of oscillations of the in-plane ($m_x$) magnetization component under the point contact(s), thus simulating the spectrum of a signal arising due to the GMR effect. Simulations for the current interval 0.5 $< I <$ 10 mA reveal two dynamical regimes: 'simple' vortex rotation around the nanocontact (for $I <$ 8 mA) and the high-current regime, where the rotating vortex generates vortex-antivortex (V-AV) pairs ($I >$ 8 mA).

From Fig. \ref{SingleVortex_Power}(c) it can be seen that in the first regime the frequency depends on the current very weakly: the total frequency shift is less than 100 MHz in the current region from 1 to 8 mA, whereas the frequency is nearly independent on current for $I > 4$ mA. This behavior is in a strong contrast with results reported for the in-plane magnetized point contacts systems \cite{Kaka_Nat_2005, Mancoff_Nat_2005}, where the frequency shift is about 100 MHz/mA. This contrast is due to the fully different nature of magnetization auto-oscillations studied in \cite{Kaka_Nat_2005, Mancoff_Nat_2005}, where most probably a nearly uniform magnetization precession (within the point contact area) around the in-plane external field was observed. In addition, due to this circumstance the oscillation frequency itself found in \cite{Kaka_Nat_2005, Mancoff_Nat_2005} is more than an order of magnitude higher than for systems studied here. 

This our result qualitatively agrees with simulation data reported by other groups, which have also studied the dynamics of vortex STNOs in confined nanoelements  induced by a spin-polarized current injected via a point contact. In particular, similarly weak $f(I)$-dependencies were found in \cite{Mistral_PRL_2008} for a SPC-induced vortex precession inside a nanodisk and in \cite{Petit-Watelot_NatPhys_2012} for a vortex STNO in a nanosquare. Quantitative comparison of our data with these results is complicated, because in \cite{Mistral_PRL_2008, Petit-Watelot_NatPhys_2012} much larger nanoelements - with lateral sizes around 1000 nm - were studied.

\begin{figure}[]
\includegraphics[width=70mm]{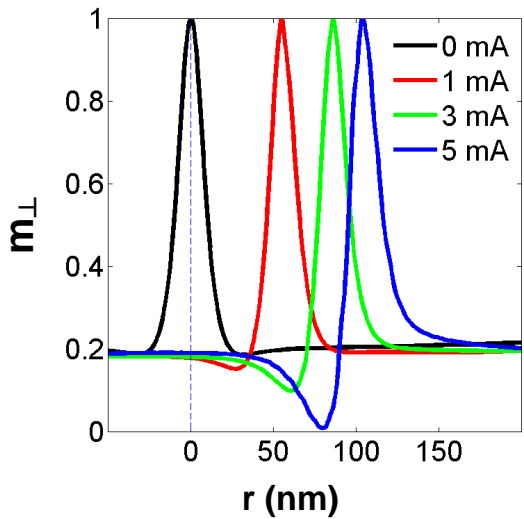}
\caption{
Vortex profiles ($m_{\perp}$ along the line connecting the vortex core and the point contact centers) for various current strengths demonstrate strong vortex deformations for large currents.
\label{VortProfiles_IncrCurr}}
\end{figure}

The weak frequency dependence on the current strength in the first regime poses the question where the additional energy pumped into the system with increasing current is spent. Analysis of simulation data shows that this energy is spent, first, to increase the radius of the vortex orbit, and second, for the dynamical deformation of the vortex structure. Both effects are clearly demonstrated in Fig. \ref{VortProfiles_IncrCurr}, where snapshots of vortex profiles (magnetization component perpendicular to the element plane) along the line between the point contact and vortex core centers are shown. Note the 'negative' peak on these $m_{\perp}(r)$ dependencies, which becomes stronger for larger currents; this vortex deformation is the necessary prerequisite for the formation of a V-AV pair \cite{Hertel_PRL_2006}, which occurs when the current is increased further.

In the high current regime, the vortex deformation exceeds the critical threshold, so that a V-AV pair is generated; in the case of the field-driven vortex dynamics this process has been first analyzed in detail in \cite{Hertel_PRL_2006}. This V-AV pair annihilates very quickly again, emitting a spin-wave burst, which is the dominant mechanism of the energy dissipation in the high-current regime. For the ST-induced vortex dynamics, the process of creation-annihilation of a V-AV pair has been described in \cite{Petit-Watelot_NatPhys_2012}, where a perpendicular external field ($H_{\perp} = 130$ Oe) much smaller than our $H_{\perp}$ was used. 

In our system the corresponding process is more complicated due to the much larger magnitude of $\bf{H}_{\perp}$. Such a large field strongly favors a core polarity along the field direction, so that the following process takes place. At the first stage, a V-AV pair is generated near the initial vortex; the polarity of the vortex from this newly generated pair is opposite to the polarity of the initial vortex. At the second stage, the antivortex from this pair annihilates with the initial vortex, releasing the exchange energy in a burst of spin waves, so that only the vortex with the opposite polarity remains. These two stages are the same as in a standard system with a small $H_{\perp}$. 

In our system, however, the process of the vortex transformation goes further. Because the value of our $H_{\perp}$ is quite large, the vortex with the polarity opposite to $\bf{H}_{\perp}$ is unstable. For this reason the third transformation stage takes place: namely, the second V-AV pair is generated near the new vortex, whereby the polarity of the vortex in this second V-AV pair coincides with the polarity of the initial vortex. Finally, the antivortex from the second V-AV pair annihilates with the vortex remained after the second stage (also emitting a spin wave burst), so that only the vortex with the same polarity as the initial one 'survives'. Then it starts to precess in the same direction as the initial vortex. A movie demonstrating this highly interesting and complicated process can be found in Supplemental Materials to this paper. A snap-shot of the out-of-plane $m$-component immediately after the annihilation of the second V-AV pair (Fig. \ref{SingleVortex_Power}(b)) shows the spin waves generated by this annihilation.

\section{III. SYNCHRONIZATION OF TWO ST-DRIVEN VORTEX NANOOSCILLATORS}

In this section we present the detailed analysis of synchronization in a systems of two nanooscillators. For this synchronization the low-current regime is obviously more suitable. As we have pointed out above, in this regime the vortex rotation frequency $f$ depends on the current strength $I$ very weakly (Fig. \ref{SingleVortex_Power}(c)). Such a weak dependence $f(I)$ is a very good prerequisite for the synchronization of STNOs in a broad current range.

\begin{figure}[]
\includegraphics[width=90mm]{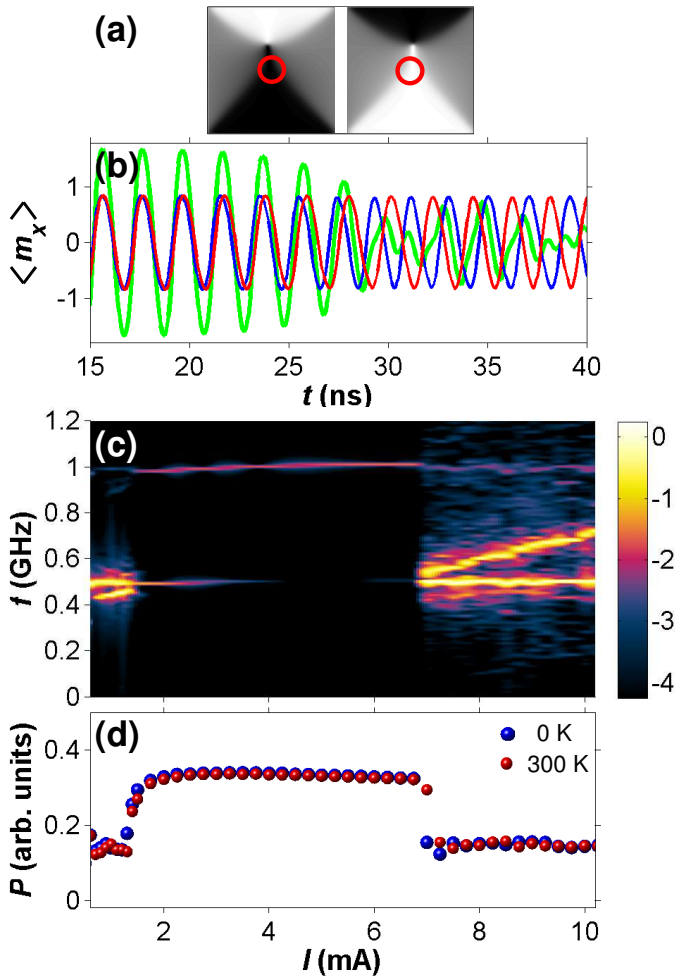}
\caption{(color online) Magnetization dynamics of two dipolar-coupled STNOs: (a) - snapshot of a magnetization configuration in a synchronized regime (note opposite oscillation phases under the contacts); (b) time dependencies of the average magnetization under the 1st ($m^x_1(t)$, red line) and 2nd ($m^x_2(t)$, blue line) contacts and the their sum ($m^x_{\rm tot}(t) = m^x_1(t)+m^x_2(t)$, green line); (c) - spectral amplitude of the total magnetization $m^x_{\rm tot}$; (d) oscillation power of the {\it difference} $m^x_{\rm diff}(t) = m^x_1(t)-m^x_2(t)$, showing the large power increase in the synchronized regime if the phase shift of $\pi$ is introduced between the ac-currents generated by the process. It can be clearly seen that when thermal fluctuations are included, the dependence $P(I)$ (panel (d)) nearly does not change.
\label{2Vort_DipCoupl_Power}}
\end{figure}

\begin{figure}[]
\includegraphics[width=90mm]{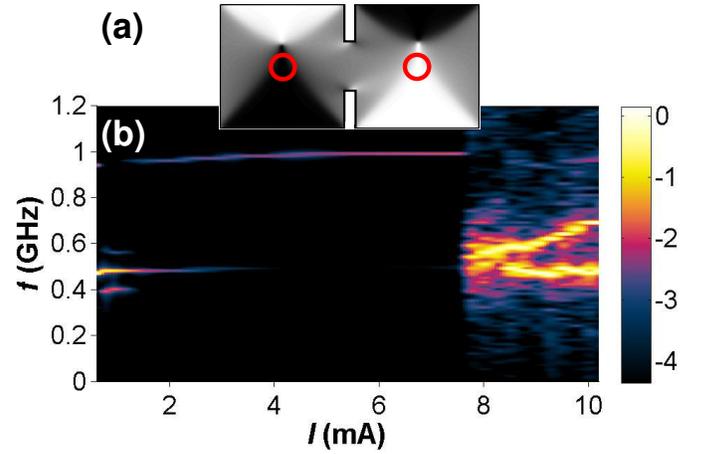}
\caption{(color online) Magnetization dynamics of the two STNOs connected by a 'bridge' (partial exchange coupling), shown in same way as in Fig. \ref{2Vort_DipCoupl_Power}(c). STNOs are synchronized almost in the whole current region, where their steady-state gyration is observed.
\label{2Vort_PartExch_Power}}
\end{figure}

\begin{figure}[]
\includegraphics[width=70mm]{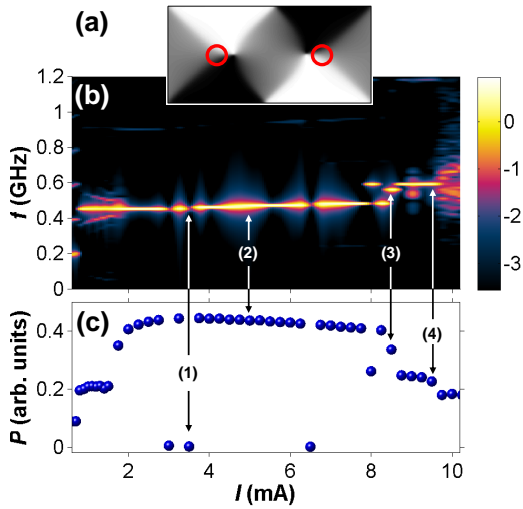}
\caption{(color online) Magnetization dynamics of the two STNOs with a full exchange coupling, presented in same way as in Fig. \ref{2Vort_DipCoupl_Power}(c). Vortex gyration is out-of-phase almost for all current values, leading to in-phase magnetization oscillations under the contacts. Numbers within double-sided arrows correspond to numbers of panels in Fig. \ref{2Vort_FullExch_DiffSynch}, where the snapshots of magnetization dynamics for corresponding currents are displayed.
\label{2Vort_FullExch_Power}}
\end{figure}

In this study, we investigate three kinds of systems with different STNO couplings: (i) dipolar coupling only: adjacent nanoelements (nanodisks and nanosquares)  separated by a gap having the width $d = 50$ nm as shown in Fig. \ref{2Vort_DipCoupl_Power}(a) for a system of two nanosquares; (ii) dipolar and partial exchange couplings: nanosquares are partially connected by the bridge with the same thickness and the width $l = 200$ nm (Fig. \ref{2Vort_PartExch_Power}(a)); and (iii) dipolar and full exchange couplings, realized as a system of two nanosquares fully connected along their sides (Fig. \ref{2Vort_FullExch_Power}(a)), so that they actually form a rectangular nanoelement with the side ratio $\approx 2:1$. For all three systems we have studied their dynamic behavior for currents through the first nanocontact in the region $0.5 < I_1 <  10$ mA, and the constant second current $I_2 = 5$ mA. Simulations for each value of $I_1$ were performed independently.

The choices of the current and spin polarization directions for these systems require a special discussion. The current direction defines the rotation sense of its Oersted field and thus - the vortex chirality (polarities of all vortices are identical due to the presence of $\bf{H}_{\rm perp}$). If the current directions are the same for both contacts, then the sign of the out-of-plane spin polarization projections $P_{\rm perp}$ is also the same. In this situation the gyration sense of both vortices coincide, providing favorable conditions for the STNOs synchronization.

However, if the rotation senses of equilibrium domain patterns are the same in both nanosquares, an antivortex is formed between the vortices in partially and fully exchange-coupled systems (ii) and (iii). Dynamics of such an antivortex can not be satisfactorily controlled by external conditions like current and external field strength even in the low-current regime. The reason is that this antivortex annihilates unavoidably with one of the oscillating vortices when the V-AV distance becomes sufficiently small, thus destroying one of STNOs. For these reasons we propose to use currents with {\it opposite} directions for adjacent point contacts, so that the chiralities of vortices created by corresponding Oersted fields are also opposite; for this configuration no antivortex is formed in exchange-coupled systems. In this case the spin polarization directions would also be opposite, so that the gyration senses of adjacent vortices remain the same.

{\it Nanoelements with the dipolar coupling only}. Results for the case (i) of dipolar-coupled nanoelements are summarized in Fig. \ref{2Vort_DipCoupl_Power}.  After the in-plane field pulse expels the vortices out of contact areas,  magnetizations within these areas rotate 'in-phase' for several tens of nanoseconds, but afterwards a transition to an 'out-of-phase' rotation occurs (Fig. \ref{2Vort_DipCoupl_Power}(b). In the steady precession regime both vortices rotate clockwise around the contacts. Spectral amplitude of the sum $m^x_{\rm tot}(t) = m^x_1(t)+m^x_2(t)$ of magnetization components under the contacts shown in Fig. \ref{2Vort_DipCoupl_Power}(c) clearly shows that synchronization of two STNOs occurs nearly in the entire current range of the regular vortex precession (i.e., where no V-AV pairs are generated) $1.5 < I_1 < 7$ mA. In this interval, the total power at the fundamental frequency $f = 0.50$ GHz almost disappears, thus signaling the nearly perfect out-of-phase synchronization of two oscillators: although the vortex cores precess in-phase (Fig. \ref{2Vort_DipCoupl_Power}(a)), magnetic moments under the contacts rotate with the phase shift $\Delta\phi = \pi$ due to the opposite vortex chiralities, so that $m^x_{\rm tot} \approx 0$. 

In an experimental system, in order to obtain the maximal output power, one would either perform the phase shift of one of the signals by $\pi$, or use the underlayers with the opposite orientations of the in-plane magnetization projections; in both approaches the GMR signals from the two contacts would become in-phase. In this case one would obtain the total oscillation power $P(I_1)$ shown in Fig. \ref{2Vort_DipCoupl_Power}(d), where we have plotted the power of $m^x_{\rm diff}(t) = m^x_1(t)+m^x_2(t) \cdot \exp(i\pi) = m^x_1(t)-m^x_2(t)$. As it should be, we observe a strong enhancement of $P$ in the synchronized regime. The power enhancement compared to the non-synchronized regime is even somewhat more than two times (the maximum power gain expected for two ideally synchronized oscillators), because for low and high currents, where the oscillators are not synchronized, the magnetization under contacts is not fully coherent.

{\it Nanoelements with the dipolar and partial exchange coupling.} For the second system, where nanosquares are connected via a bridge, STNOs interact both via dipolar and exchange interactions. Because we could avoid the formation of an antivortex between the vortices, we have achieved a stable steady-state precession of vortices and their synchronization (Fig. \ref{2Vort_PartExch_Power}(a)) also for this system. Synchronized magnetization oscillations under the contacts are here also out-of-phase. 

Due to higher coupling energy for this system (see the discussion below), synchronization starts to establish itself already for the smallest current where the vortex precession is induced (see Fig. \ref{2Vort_PartExch_Power}(b)). The maximal current, for which the synchronization is observed, is somewhat larger than for the pure dipolar coupling of nanosquares. Due to the presence of the 'bridge' between the nanosquares, we expect the coupling energy here to be significantly larger than for the case of nanoelements with the dipolar coupling only. We shall return to this question later in the subsection devoted to the evaluation and comparison of coupling energies in various systems.

\begin{figure}[]
\includegraphics[width=70mm]{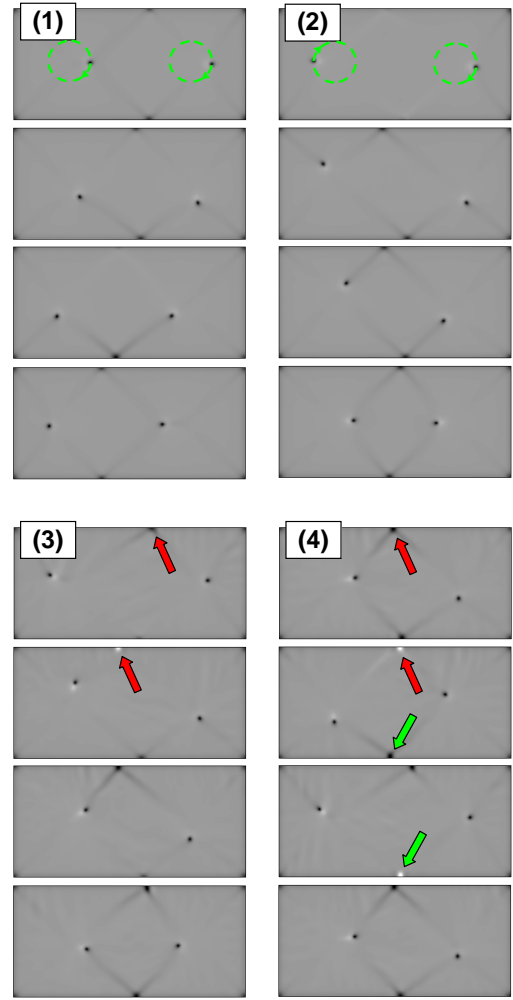}
\caption{
Snapshots of the $m_{\perp}$-projection for the four different synchronized states of two STNOs with a full exchange coupling (see Fig. \ref{2Vort_FullExch_Power}): (1) - in-phase oscillations of vortices for $I$ = 3.5 mA, (2) - out-of-phase oscillations of vortices for $I$ = 5.0 mA, (3) - nearly in-phase oscillations of vortices for $I$ = 8.5 mA accompanied by the polarity reversal of the half-antivortex at the upper system border (shown with red arrows), (4) oscillations of vortices for $I$ = 9.2 mA accompanied by the polarity reversal of both half-antivortices - at the upper and lower ((shown with green arrows) system borders, whereby the phase difference between the vortex precessions is neither 0 nor $\pi$.
\label{2Vort_FullExch_DiffSynch}}
\end{figure}

{\it Nanoelements with the dipolar and full exchange coupling.} The third system - fully connected nanosquares (so that they actually form a single rectangle) - demonstrates a qualitatively different type of synchronization as shown in Fig. \ref{2Vort_FullExch_Power}(a): the strong exchange interaction between STNOs leads for most currents to out-of-phase vortex core oscillations, so that magnetization oscillations under the contacts are {\it in-phase}. Snapshots of the $m_{\perp}$-component, where the out-of-phase precession of vortex core is clearly visible, are shown in Fig. \ref{2Vort_FullExch_DiffSynch}, panel (2). However, the out-of-phase synchronized state is also possible, as it can be seen from the spectral power 'dips' at $I_1 = 3.0$, 3.5 and 6.5 mA in Fig. \ref{2Vort_FullExch_Power}(a,b). Snapshots of $m_{\perp}$ for this case - in-phase oscillations of vortex cores - are presented in Fig. \ref{2Vort_FullExch_DiffSynch}, panel (1). 

Keeping in mind a comparison to possible experiments, we remind here once more that simulations are performed for each current value independently, so that for each current the specific synchronization state establishes itself independently on the states achieved for 'neighboring' currents. We also note, that the existence of different synchronized states for one and the same current in STNO systems has been reported by one of us for a very different design of two synchronized out-of-plane STNOs on a nanowire \cite{Berkov_PRB2013}.

Synchronization in the high-current regime ($I > 8$ mA), where also several  different synchronized states are possible (Fig. \ref{2Vort_FullExch_Power}(b)) deserves a special attention. In a nanorectangle with two vortices of opposite chiralities, both equilibrium and dynamical magnetization states contain two half-antivortices approximately in the middle of the two long (horizontal) sides of the rectangle, where the 90-degree domain walls touch these sides (see, e.g., Fig. \ref{2Vort_FullExch_Power}(a)). For sufficiently high currents the energy supplied to the system by the spin-polarized current is so large that the polarity switching of one (Fig. \ref{2Vort_FullExch_DiffSynch}, panel (3)) or both (Fig. \ref{2Vort_FullExch_DiffSynch}, panel (4)) these antivortices occurs. Corresponding points are marked in Fig.\ref{2Vort_FullExch_DiffSynch} (panels (3) and (4) by arrows showing the location of half-antivortices which polarity switches during the magnetization oscillations. The frequencies of oscillations involving such polarity change(s) is different from that for the 'normal' vortex precession, which causes frequency and power jumps seen in Fig. \ref{2Vort_FullExch_Power}(b,c).

{\it Coupling energy of two vortex STNOs.} One of the most important parameters of a synchronized system is the coupling energy between the oscillators $E_{\rm coup}$. This quantity defines not only the current range where synchronization occurs and the maximal allowed frequency mismatch between single oscillators \cite{Slavin_PRB_2006, ChenXi_PRB_2009,  Belanovsky_PRB_2012}, but also the synchronization stability against thermal fluctuations. For some simple cases there exist analytical and semianalytical methods for the estimation of $E_{\rm coup}$ \cite{Slavin_PRB_2006, Belanovsky_PRB_2012}. However, for complex systems like dipolar- and exchange-coupled vortices reliable analytical methods to compute this quantity are lacking. 

For this reason we suggest here a purely numerical general method for calculation of $E_{\rm coup}$. First we compute the {\it mean} total system energy $\langle E_{\rm tot} \rangle$, averaging the total energy in the dynamical (synchronized) regime (which oscillates slightly in time) $E_{\rm tot}(t)$ over several oscillation periods. Subtracting from this quantity the equilibrium system energy $E_{\rm eq}$ at $T = 0$ (for the state where the Oersted field is present but the spin torque is still switched off), we obtain the dynamical contribution to the total energy in the synchronized regime $E^{\rm dyn}_{\rm sync}$, pumped into the system by the spin torque: $E^{\rm dyn}_{\rm sync} = \langle E_{\rm tot} \rangle - E_{\rm eq}$ (this energies depends, of course, on both current strengths $I_1$ and $I_2$). Then we perform the same procedure for a single STNO, obtaining the dynamical additives $E^{\rm dyn}_{1(2)}$ for oscillators 1 and 2 in the case where there is no interaction between them in dependence on the current flowing through a single STNO. The difference between the dynamical additives for the system of synchronized STNOs $E^{\rm dyn}_{\rm sync}$ and the sum of $E^{\rm dyn}_1+E^{\rm dyn}_2$ of two STNOs operating separately (at the same current strengths $I_1$ and $I_2$ as the synchronized system) provides a dynamical coupling energy: $E_{\rm coup} = E^{\rm dyn}_{\rm sync} - (E^{\rm dyn}_1 + E^{\rm dyn}_2)$.

\begin{figure}
\includegraphics[width=70mm]{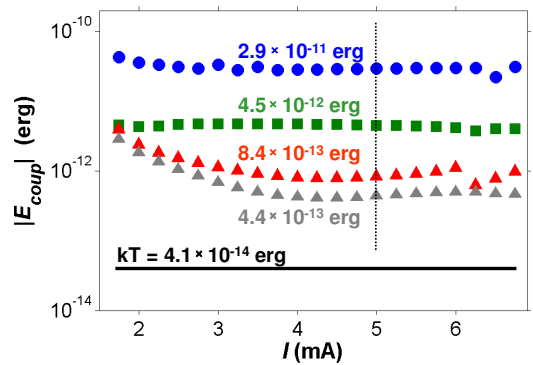}
\caption{(color online) Coupling energy vs current strength in the 1st nanocontact $I_1$ (for the same $I_2 = 5$ mA) for systems of two vortex STNOs: grey triangles - nanodisks with dipolar coupling only, red triangles - nanosquares with dipolar coupling only (Fig. \ref{2Vort_DipCoupl_Power}(a)), squares - 'bridge' configuration (Fig. \ref{2Vort_PartExch_Power}(a)), circles - full exchange and dipolar coupling (Fig. \ref{2Vort_FullExch_Power}(a)). 
\label{CouplEnerg_VarSys}}
\end{figure}

Results of this calculation are presented in Fig. \ref{CouplEnerg_VarSys}, where the plots of the {\it absolute values} of $E_{\rm coup}(I_1)$ are shown (note the logarithmic scale of the energy axis). The numbers on the plot are the coupling energies for $I_1 = 5$ mA ($I_2 = 5$ mA, as always). First we note, that even for the system with only the dipolar interaction between STNOs the lowest coupling energy (obtained for $I_1 > 4$ mA) is roughly twenty times larger than $kT$. Hence the effect of thermal fluctuations on the dynamics of this system should be negligible. We have verified this expectation by simulations of this system including thermal fluctuations at $T = 300$ K. The spectral amplitude map $A(f, I_1)$ for this temperature (not shown) is very similar to that presented in Fig. \ref{2Vort_DipCoupl_Power}(c) for $T = 0$. Comparison of total oscillation powers for $T = 0$ and $T = 300$ K (Fig. \ref{2Vort_DipCoupl_Power}(d)) demonstrates that thermal fluctuations do not really affect the synchronization dynamics. 

For the fully exchange-coupled system of nanosquares the relation of a typical coupling energy (shown in Fig. \ref{CouplEnerg_VarSys} with blue solid circles) to the thermal energy is $E_{\rm coup}/kT \approx 2.9 \times 10^{-11}/4.1 \times 10^{-14} \approx 700$, so that in this system thermal fluctuations should play no role whatsoever.

To present an example how the coupling energy depends on the element shape, we have performed analogous simulations for the system of two {\it circular} nanoelements (having the same material parameters), with diameters $D = 500$ nm and the smallest edge-to-edge separation $d = 50$ nm, as for the system of nanosquares. The spectral amplitude map $A(f, I_1)$ for the system of such nanodisks was virtually indistinguishable from that for a system of nanosquares (Fig. \ref{2Vort_DipCoupl_Power}(c)), except that all basic frequencies were $\approx 50$ MHz larger and the 2nd harmonics frequencies - correspondingly $\approx 100$ MHz higher. The coupling energy for the system of nanodisks is shown in Fig. \ref{CouplEnerg_VarSys} with grey triangles. This energy is always smaller than for a pair of vortex STNOs in nanosquares, what can be explained by the smaller total volume of nanodisks, which leads to smaller total magnetic 'volume charges', partly responsible for the coupling energy of STNOs in our case.

We also note, that our coupling energy for a system of two nanodisks $E_{\rm coup} = 4.4 \times 10^{-13}$ erg has the same order of magnitude as $E_{\rm coup} = 2.3 \times 10^{-13}$ erg obtained in \cite{Belanovsky_PRB_2012} also for two vortex-based STNOs in nanodisks with the edge-to-edge distance 50 nm. However, nanodisks simulated in \cite{Belanovsky_PRB_2012} have the diameters 200 nm and thicknesses 10 nm so that the quantitative comparison between our results is not feasible.

\begin{figure*}[]
\includegraphics[width=140mm]{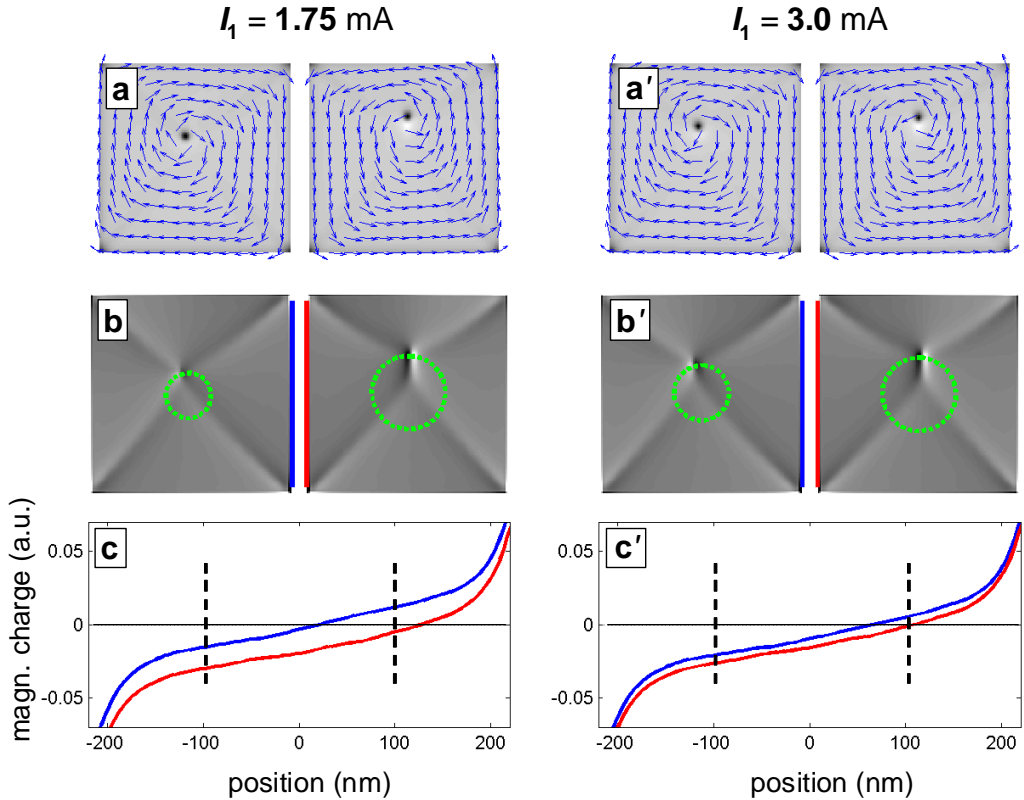}
\caption{(color online) To the explanation of the coupling energy dependence on the current $I_1$: (a) and (a') - snapshots of the magnetization configurations as arrows; (b) and (b') - volume charges; (c) and (c') - surfaces charges along the adjacent nanosquare surfaces (marked on (b) and (b') with blue and red lines) for two different currents $I_1 = 1.75$ mA (left column of images) and $I_1 = 3.0$ mA (right column of images). For the configuration with the 'bridge' (see Fig. \ref{2Vort_PartExch_Power}) surface charges between vertical dashed lines shown in parts (c) and (c') would not exist.
\label{Magn_charges_Dip_Coupl}}
\end{figure*}

The most interesting feature of the coupling energy $E_{\rm coup}(I_1)$ as the function of the current $I_1$ flowing through the $1^{\rm st}$ nanocontact is the rapid decrease of this energy with increasing $I_1$ for systems with the dipolar coupling only (red and grey triangles in Fig. \ref{CouplEnerg_VarSys}). At the same time, $E_{\rm coup}$ depends on the current $I_1$ relatively weakly for the pairs of nanosquares which are partially of fully connected (squares and circles in Fig. \ref{CouplEnerg_VarSys}).

To understand this behavior, we have studied the distribution of 'magnetic charges' in our system. Volume $\rho({\bf r})$ and surface $\sigma({\bf r})$ densities of these 'charges' were computed numerically as the divergence of the magnetization field $\rho = - \nabla M$ and the magnetization component perpendicular to the corresponding surface $\sigma({\bf r}) = M_{\perp}$ from the magnetization configuration obtained in simulations. 

Two examples of charge distributions obtained this way for the pair of separated nanosquares at different current values $I_1$ are given in Fig. \ref{Magn_charges_Dip_Coupl}. Here we show - for currents $I_1 = 1.75$ mA and  $I_1 = 3.0$ mA - the magnetization configurations (images (a) and (${\rm a}^\prime$))), volume charges within the whole system (grey scale maps (b) and (${\rm b}^\prime$)), and surface charges along the adjacent edge surfaces of the two nanosquares (graphs (c) and (${\rm c}^\prime$)). On the images (b) and (${\rm b}^\prime$), the vortex orbit in each nanosquare is drawn with the green dotted line. In Fig. \ref{Magn_charges_Dip_Coupl}, surface charges shown with red and blue solid lines on the graphs (c) and (${\rm c}^\prime$)) are calculated along the adjacent nanosquare edges marked on the images (b) and (${\rm b}^\prime$)) in red and blue, correspondingly.

From the comparison of the surface charges on adjacent vertical nanosquare edges for two different currents (graphs (c) and (${\rm c}^\prime$)) the following trend can be clearly seen. In the left nanosquare, for the low current $I_1 = 1.75$ mA the surface charge $\sigma(y)$ is a nearly odd function of the coordinate $y$ along the vertical edge (if the coordinate origin is placed in the edge center), i.e. it changes sign approximately in the middle of this edge. In the right nanosquare - due to the much higher current $I_2 = 5$ mA - the radius of the vortex orbit is significantly larger, and the surface charge $\sigma(y)$ retains the same sign almost along the entire vertical left edge of this nanoelement. For this reason, the total magnetostatic interaction energy between surface charges on vertical edges marked in blue and red on the panel (b), is relatively small. 

When the current through the left STNO increases to $I_1 = 3.0$ mA (right column of images in Fig. \ref{Magn_charges_Dip_Coupl}), the functional dependence of the surface charge along the right edge of the left square $\sigma_l(y)$ becomes very similar to the corresponding function $\sigma_r(y)$ for the left edge of the right nanosquare, as shown in the graph (${\rm c}^\prime$). This means that the magnetodipolar interaction energy between these two surfaces - marked in blue and red on panels (b) and (${\rm b}^\prime$) becomes much larger. In addition, from the inspection of plots in Fig. \ref{Magn_charges_Dip_Coupl}${\rm c}^\prime$ it is clear that this energy is definitely positive - 'blue' and 'red' surface charges have the same signs almost everywhere along the adjacent edges of neighboring nanosquares. 

Hence from the comparison of the charge configurations for the smaller (1.75 mA) and larger (3.0 mA) currents $I_1$ flowing through the left nanocontact, it follows that for the larger value of $I_1$ the magnetostatic interaction energy between the surface charges on adjacent vertical edges of the nanoelements is much higher - and positive - for the larger current. Taking into account that the total coupling energy $E_{\rm coup}$ is {\it negative}, the {\it positive} contribution to this energy which becomes {\it larger} with increasing $I_1$, leads to the decrease of the absolute value of $E_{\rm coup}$ for separated nanoelements as seen in Fig. \ref{CouplEnerg_VarSys}. We also point out, that the contribution under discussion comes from the strong interaction between magnetic charges located on closely placed surfaces, so that it is one of the major contributions to the total coupling energy. 

For the system of partially coupled nanosquares (Fig. \ref{2Vort_PartExch_Power}), the bridge between the nanoelements in the middle of the structure eliminates the surface charges coming from the central region of adjacent vertical edges of nanosquares. This reagion is marked in Fig. \ref{Magn_charges_Dip_Coupl} with vertical dashed lines on panels (c) and (${\rm c}^\prime$): the surface charges in-between these lines do not exist for the 'bridged' configuration, so that their positive magnetodipolar interaction energy does not contribute to the total (negative) coupling energy. For this reason the coupling energy for the 'bridged' configuration depends on the current $I_1$ only weakly and for large current values becomes significantly larger than $E_{\rm coup}$ for the fully separated nanosquares.

An additional observation supporting this argumentation comes from the comparison of  coupling energies of the 'bridged' (Fig. \ref{2Vort_PartExch_Power}) and fully connected (Fig. \ref{2Vort_FullExch_Power}) configurations. The coupling energy for the latter system is much higher (compare the dependency marked with blue circles to that marked by green squares in Fig. \ref{CouplEnerg_VarSys}). This difference is most probably due to the complete absence of surface charges on adjacent closely placed vertical edges in the fully connected system (because these edges are now absent). From the plots in Fig. \ref{Magn_charges_Dip_Coupl} (panels (c) and (${\rm c}^\prime$)) it is evident that the surface charges on the adjacent vertical edges near the nanosquare corners have always the same sign, so that their magnetostatic interaction energy is large and positive. Elimination of this contribution to the (negative) coupling energy in the fully connected configuration should increase this energy, as confirmed by Fig. \ref{CouplEnerg_VarSys}.

Here we would like also to note that a related problem - coupling of the vortex resonance modes in closely placed nanoelements - was studied in \cite{Jain_PRB2012}, where the system of two overlapping nanodisks was analyzed. The major effect found in \cite{Jain_PRB2012} - change of the resonant frequency when the disk overlapping and/or the external field have been varied - was due to the change of the magnetic volume charges when two vortices in the overlapping nanodisks approached each other. The main difference between this study and our work (not to mention that spin torque effects have not been studied in \cite{Jain_PRB2012}) is the following: due to another shape studied in our paper - squares instead of disks - our configuration of magnetic charges is qualitatively different from that calculated in \cite{Jain_PRB2012}, so the main effect in our case comes from the change of the surface charge configuration.

Concluding this discussion, we point out the following very important circumstance: according to the consideration presented above, the overwhelming contribution to the coupling energy in our systems comes from the magnetodipolar interaction between magnetization configurations of different STNOs - even for the partially and fully connected nanosquares, where the interaction via the spin waves could also be an option. This was confirmed by the analysis of the spatial maps of the oscillation power, where magnetization oscillations in the 'bridge' region between the nanosquares and in the middle of the fully connected structure were negligibly small. This behavior is in strong contrast with system of synchronized STNOs studied experimentally in, e.g., \cite{Kaka_Nat_2005, Mancoff_Nat_2005}, for which it could be shown \cite{Slavin_PRB_2006} that the coupling via the spin waves is the dominant mechanism. In most cases investigated numerically \cite{Berkov_PRB2013, ChenXi_PRB_2009} the coupling via the spin waves emission and absorption also dominated the establishing of the synchronized regime.

The dominating role of the magnetodipolar interaction by synchronization of vortex-based STNOs should have, in particular, far-reaching implications for developing of spintronic devices where the enhancement of the output microwave power using several synchronized STNOs is aimed. Namely, due to the very high speed of electromagnetic waves (which is virtually infinite on the time scale of interest in such devices), there should be no phase shift between the synchronized STNOs, if the dominant synchronization mechanism is the magnetodipolar interaction. For this reason the output power  in such systems should scale as $~ N^2$ with the number of STNOs no matter how many oscillators are synchronized. This very favorable scaling can not be obtained, in general, in system of spin-wave-synchronized STNOs: the finite (and not even very large) velocity of spin waves always leads to a noticeable phase shift between adjacent oscillators, so that starting from a certain number of STNOs the total output GMR power ceases to increase. We shall come back to this point below by analyzing the synchronization behavior of large vortex-based STNO arrays.

\section{IV. SYNCHRONIZATION OF A LARGE NUMBER OF VORTEX-BASED NANOOSCILLATORS: 1D ARRAYS}

\begin{figure*}[]
\centering
\includegraphics[width=150mm]{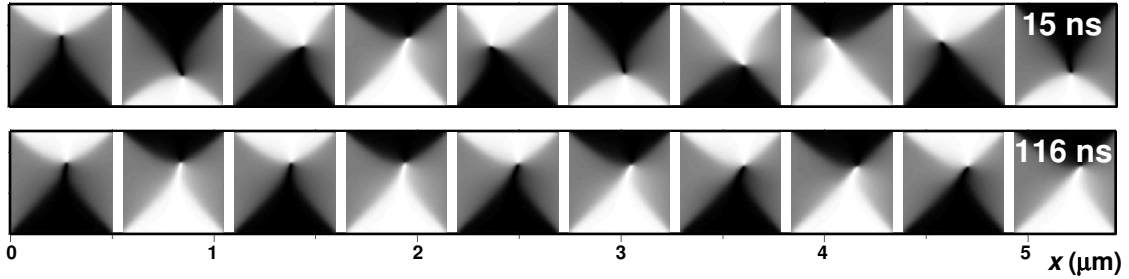}
\caption{Snapshots of the in-plane magnetization distributions for 10 nanooscillators in the transient ($t = 15$ ns, upper panel) and synchronized ($t = 116$ ns, lower panel) regimes.
\label{1x10_DipCoupl_Snapshots}}
\end{figure*}

In this Section we show that our design allows to obtain the easy and robust synchronization of an {\it arbitrary} number of vortex-based STNOs. To demonstrate this feature, we have first simulated several systems consisting of many STNOs arranged as a 1D array (chain). 

{\it Chain of STNOs coupled via the dipolar interaction only}. Here we consider an example of 10 dipolarly coupled oscillators placed as shown in Fig.\ref{1x10_DipCoupl_Snapshots} (the total system size is $5430 \times 500$ nm$^2$). Due to the large number of oscillators, for this system it is obviously impossible to use the 'current ramping' protocol employed for two STNOs, i.e. keeping $I_2 = {\rm Const}$ and varying $I_1$: an attempt to use such a procedure for a system of $N$ STNOs would result in a 'phase diagram' in a ($N-1$)-dimensional space. For this reason we have simulated a system where {\it total} currents (not the current densities) flowing through all contacts are equal: $I_i = I$ for $i=1,...,N$, where the current value $I$ was varied from 0.5 to 7 mA. 

Using this simulation protocol, we have addressed the following very important problem: how stable is the synchronization with respect to experimentally unavoidable random deviations of the contact diameters from their nominal values. Importance of this question becomes evident from descriptions of experiments employing the point contact setups: whereas the dimensions of nanodisks and nanosquares with lateral sizes $\sim 1$ $\mu$ can be controlled with the accuracy of 10 nm (i.e. with 1\% relative precision), manufacturing of nanocontacts with a prescribed diameter remains a challenging experimental task - see e.g. \cite{Rippard_PRL_2004, Yanson_PRL_2005, Mistral_PRL_2008, Bonetti_PRL_2010} etc. As a consequence of this difficulty, point contact diameters determined from independent experimental methods \cite{Rippard_PRL_2004, Mistral_PRL_2008} or by adjustment the signal frequency using micromagnetic simulations \cite{Bonetti_PRL_2010} can differ by more than 50\% from intended values (for this reason, one usually speaks about 'nominal diameters' of such contacts). Hence, numerical studies of synchronization of many STNOs with significant dispersion of the point contact sizes are crucially important from the experimental point of view.

\begin{figure}[]
\includegraphics[width=70mm]{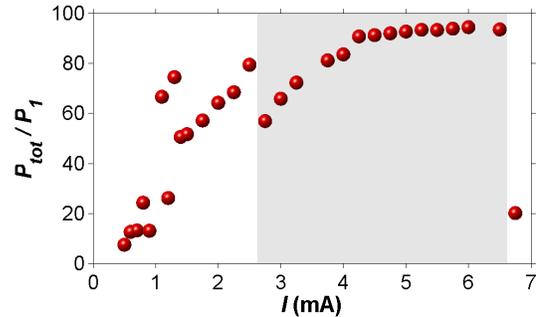}
\caption{
Total oscillation power (normalized to the oscillation power of one STNO) of the alternating sum of the in-plane components inside the point contacts $m^x_{\rm alt}(t) = \sum_i (-1)^{i-1}m^x_i(t)$, which takes into account the opposite oscillation phases of the magnetization in adjacent contacts.
\label{1x10_DipCoupl_Power}}
\end{figure}

Simulating a single STNO, we have found that the vortex gyration frequency is nearly independent on this diameter, if the total current is kept constant. The reason is, that the vortex oscillation orbit lies outside the point contact, where the Oersted field (and thus - the vortex dynamics) depends only on the {\it total} current through the contact. Hence a robust synchronization of many STNOs based on point contacts with very different diameters can be expected, as long as the currents (which can be controlled experimentally very precisely) through all contacts are the same. In our simulations we have used a system of 10 STNOs, with contacts diameters $D_i$ chosen randomly from the Gaussian distribution with the mean $D_{\rm av} = 90$ nm and dispersion $\sigma = 30$ nm (randomly generated values of $D_i$ were within the region $62 < D_i < 118$ nm). Fig. \ref{1x10_DipCoupl_Snapshots} displays in-plane magnetization distributions for two time moments: 15 ns after simulations have been started and 100 ns later, when oscillators are almost perfectly synchronized. Plot of the total oscillation power (normalized to the oscillation power of one STNO) vs current displayed in Fig. \ref{1x10_DipCoupl_Power} shows that an impressive power gain of nearly 100 times - close to the maximum theoretical value - can be obtained in this highly non-ideal system of point contact STNOs for currents $I_0 > 4$ mA.

The nearly $N^2$-scaling of the oscillations power achieved in this system for currents $I > 4$ mA illustrates the possibility to obtain the almost perfect in-phase synchronization of STNOs  - i.e. without any phase shift between different STNO - when the magnetodipolar interaction is the dominating mechanism (see the corresponding discussion at the end of the previous Section). The fact that power gain in this case is slightly smaller than $N^2 = 100$ is partially due to a relatively weak coupling between the oscillators and partially - due to the finite size effect (the STNOs at the ends of the chain are in somewhat different conditions than those in the chain middle). 

\begin{figure}[]
\includegraphics[width=70mm]{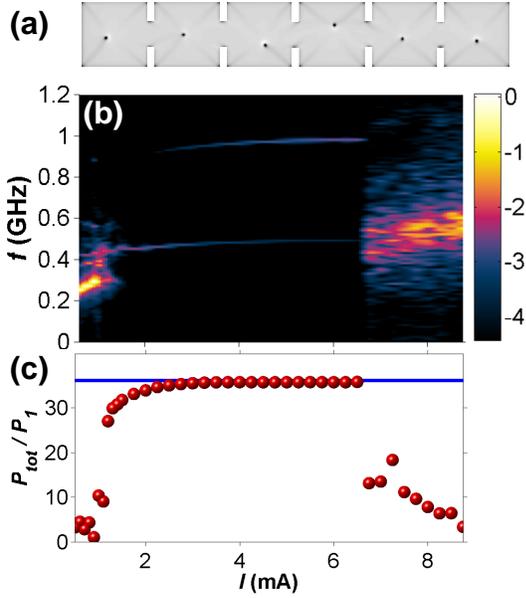}
\caption{
(color online) Magnetization dynamics of 6 STNOs connected by the 'bridges' width the same widths $l=200$ nm as in Fig. \ref{2Vort_PartExch_Power} (partial exchange coupling), shown in same way as in previous figures. The perfect synchronization is demonstrated by comparing the relation of the total power to the power of one vortex oscillator $P_{tot}/P_1$ (red circles at the bottom graph (c)) with the square of the STNOs number $N^2 = 36$ (blue horizontal line at the same plot).
\label{1x6_PartExch_Power}}
\end{figure}

{\it Chain of STNOs coupled via the dipolar and partial exchange interactions}.
Next, we have studied the STNO synchronization in a 1D array of $N=6$ partially exchange coupled nanosquares, i.e. in a chain of nanosquares connected via a bridge as shown in Fig. \ref{1x6_PartExch_Power}(a). We have used the same simulation protocol as for the array of separated nanosquares described in the previous subsection, i.e. current values flowing through all nanocontacts were identical; diameters of nanocontacts were also generated from the Gaussian random distribution with the same parameters as listed above ($D_{\rm av} = 90$ nm and $\sigma = 30$ nm) to test the synchronization stability with respect to random variation of this quantity.

Previously we have demonstrated that the coupling energy in this system is much higher than $E_{\rm coup}$ for separated STNOs (see Fig. \ref{CouplEnerg_VarSys}). In accordance with this result, the chain of partially coupled oscillators is not only synchronized almost in the whole current region of the steady-state vortex gyration (\ref{1x6_PartExch_Power}(b)), but (starting from the current $I \approx 3$ mA) the total output power is equal to its maximal possible value $P_{max} = N^2 P_1 = 36 P_1$, as it can be clearly seen from Fig. \ref{1x6_PartExch_Power}(c). For this system the steady-state oscillation regime also corresponds to the phase shift $\Delta\phi = \pi$ between magnetization oscillations under neighboring contacts. For this reason the output power was evaluated as the oscillation power of the alternating sum of the $m_x$-components inside the point contacts $m^x_{\rm alt}(t) = \sum_i (-1)^{i-1}m^x_i(t)$, analogous to that shown in Fig. \ref{1x10_DipCoupl_Power}.

\begin{figure}[]
\includegraphics[width=70mm]{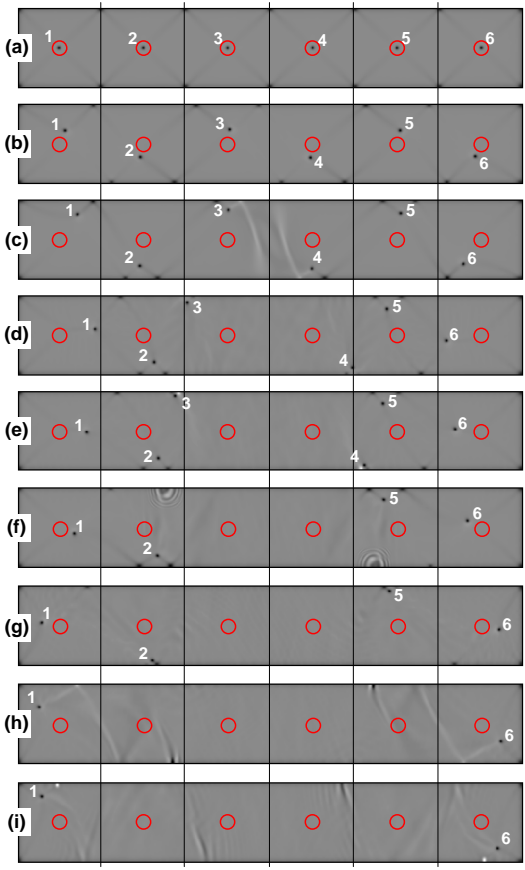}
\caption{
Snapshots of the vortex annihilation stages in a system of fully exchange coupled nanosquares on the top and bottom system borders due to the absence of the confinement from side borders of nanosquares.
\label{1x6_FullExch_Snapshots}}
\end{figure}

{\it Chain of STNOs coupled via the dipolar and full exchange interactions}. When $N$ nanosquares are connected with 'bridges' of the widths equal to the nanosquare side $a$, the resulting system becomes a nanostripe with the width $a$ and the length $\approx N \cdot a$, as shown in Fig. \ref{1x6_FullExch_Snapshots}. In this figure, red circles mark the point contacts, and thin vertical lines represent the (here imaginary) boundaries of squares out of which the nanostripe has been 'assembled'.

For such a system we could not find any regime where a steady state precession of all vortices outside the point contact - not to mention their synchronization - could be achieved. Snapshots of the transient magnetization dynamics shown in Fig. \ref{1x6_FullExch_Snapshots} demonstrate a typical time evolution of the initial vortex structure (Fig. \ref{1x6_FullExch_Snapshots}(a)). First, when the initial in-plane pulsed magnetic field is applied, the vortices are expelled out of the point contact areas (Fig. \ref{1x6_FullExch_Snapshots}(b)) and start to precess around the corresponding point contacts (Fig. \ref{1x6_FullExch_Snapshots}(c)). After a few nanoseconds the two vortices closest to the nanostripe center (vortices 3 and 4 in Fig. \ref{1x6_FullExch_Snapshots}) leave the squares where they have been initially created by the Oersted field of the corresponding current - see Fig. \ref{1x6_FullExch_Snapshots}(d). Next, these vortices arrive at the upper (vortex 3) and lower (vortex 4) horizontal stripe borders (Fig.\ref{1x6_FullExch_Snapshots}(e)) and annihilate there emitting a burst of spin waves (Fig. \ref{1x6_FullExch_Snapshots}(f)). Then the same happens to vortices 2 and 5 (Fig. \ref{1x6_FullExch_Snapshots}(g,h)), so that finally only the the first and last vortices 'survive' within a nanostripe - see Fig. \ref{1x6_FullExch_Snapshots}(i). A qualitatively similar process was observed for all values of the initial pulsed field which were large enough to expel the vortices out of the point contact areas.

The reason for this instability is obviously the insufficient spatial confinement for the motion of vortices within all squares, except the first and the last ones. The confinement provided only by the lower and upper borders of the nanostripe is not strong enough to avoid the 'runaway' of the vortices and their annihilation on the nanostripe edges. This finding makes clear that the robust steady state precession and synchronization of two vortices in the fully exchange coupled system of two nanosquares (nanorectangle with the side ratio 2:1 shown in Fig. \ref{2Vort_FullExch_Power}) is due to the fact that in this system the motion of each vortex is confined from three sides. Such a confinement is strong enough to ensure the steady gyration of vortices.

Results of this subsection mean that for the achievement of a robust synchronization in a system of many vortex STNOs, the system of nanosquares connected via 'bridges' (with the widths much shorter than the side of the nanosquare) represents an optimal compromise between the required high coupling energy (which is maximal for fully coupled nanosquares) and the stability of the vortex orbit due to the spatial confinement provided by the nanoelement borders (which is maximal for fully separated nanosquares).

\section{V. SYNCHRONIZATION OF A LARGE NUMBER OF VORTEX-BASED OSCILLATORS: A 2D ARRAY}

The system concept for the STNO synchronization suggested in the previous Section can be directly generalized to two-dimensional STNO arrays. Namely, a square lattice of nanoelements (nanosquares or nanodisks), with the point contacts placed in the centers of these elements and current directions through the contacts forming a checkerboard pattern, should support a synchronized precession of STNOs with the corresponding huge increase of the total signal power.

\begin{figure}[]
\centering
\includegraphics[width=65mm]{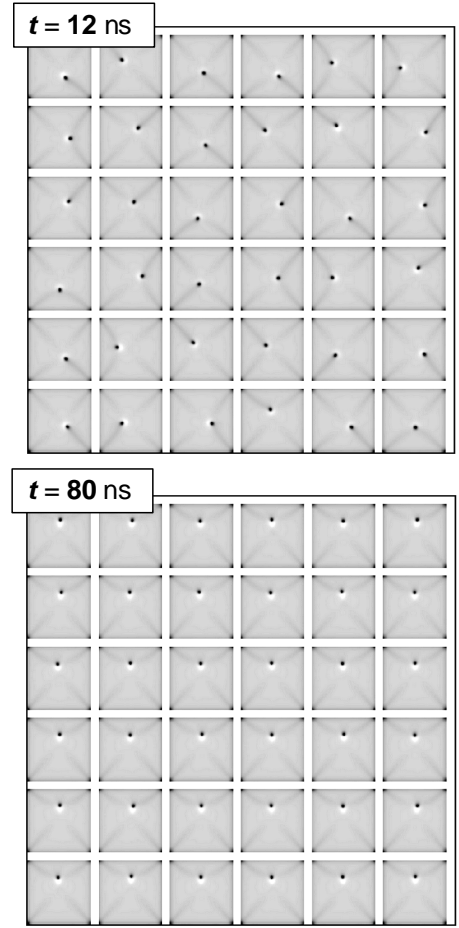}
\caption{
Snapshots of the out-of-plane magnetization projection for a 2D system of STNOs ($6 \times 6$ lattice) shortly after the current has been switched on (upper image) and after a perfect synchronization is achieved (lower image). Currents through all contacts are equal to $I = 5$ mA.
\label{6x6_latt_Snapshots}}
\end{figure}

\begin{figure}[]
\includegraphics[width=70mm]{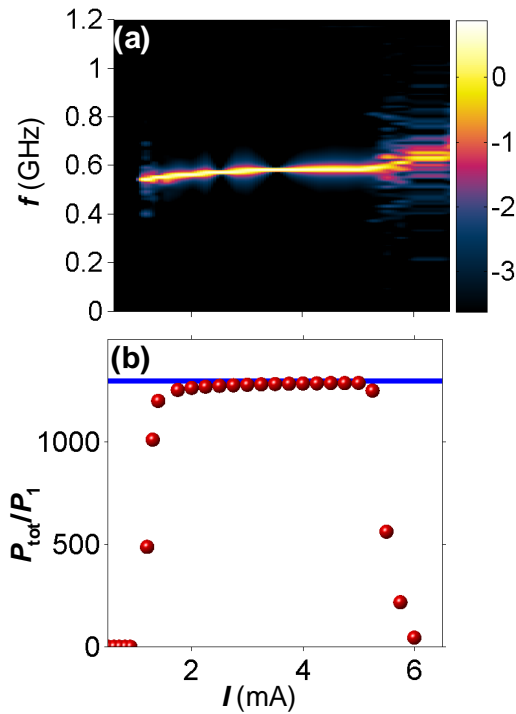}
\caption{
(color online) Magnetization dynamics of a $6 \times 6$ lattice of STNOs shown in Fig. \ref{6x6_latt_Snapshots}, presented in same way as in Fig. \ref{1x6_PartExch_Power}. In this case the perfect synchronization is also demonstrated by comparing the relation $P_{tot}/P_1$ (red circles at the bottom graph (b)) with the square of the STNOs number $36^2 = 1296$ (blue horizontal line at the same graph).
\label{6x6_latt_Power}}
\end{figure}

Simulation example of such a system consisting of $6 \times 6$ fully separated nanosquares is shown in Fig. \ref{6x6_latt_Snapshots}. Simulations were performed for periodic boundary conditions, in order to reduce the simulation time. Here we have again used point contacts with the same spread of their diameters as in Sec. IV (i.e. their diameters were generated using a Gaussian distribution with $D_{\rm av} = 90$ nm and $\sigma = 30$ nm). This dispersion of nanocontact sizes explains the chaotic initial vortex positions soon after the in-plane field pulse - used to expel the vortex out of the contact areas - was switched off (see upper panel of Fig. \ref{6x6_latt_Snapshots}). However, after $\approx 50$ ns a perfect synchronization of the vortex precessions is achieved, as shown in Fig. \ref{6x6_latt_Snapshots}, lower panel.

In this system we have observed two synchronization regimes. For small currents, the  vortices in neighboring nanoelements rotate in the opposite phases and the GMR signals coming from neighboring contacts would be in-phase. For currents larger than $\approx 1$ mA, neighboring vortices rotate in-phase and would correspondingly generate signals with the phase shift $\Delta \phi = \pi$ (this case is shown in Fig. \ref{6x6_latt_Snapshots}). However, we note that the value of the current where the changing of the synchronization regimes occurs, decreases with increasing system size, so that for a sufficiently large system only the in-phase vortex synchronization might be a stable regime.

The map of the spectral amplitude of the total magnetization oscillations under the point contacts in the $f-I$ plane is shown in Fig. \ref{6x6_latt_Power}(a). Here the oscillation power for the checkerboard-alternating 2D sum of $m_x$-components  $m^x_{\rm alt}(t) = \sum_{ij} (-1)^{i+j} m^x_{ij}(t)$ is shown, because for the majority of simulated currents (namely, for $I > 1$ mA), the magnetizations under neighboring nanocontacts oscillate with the opposite phases. 

The dependence of the total oscillation power on the current plotted in Fig. \ref{6x6_latt_Power}(b) demonstrates that the maximal possible power of $P_{tot}/P_1 = (6 \times 6)^2 = 36^2 = 1296$ is achieved already for currents only slightly higher than the critical current when the corresponding synchronized regime establishes itself. This much more efficient synchronization - as compared to the case of a 1D chain of separated nanosquares - is explained by the larger number of nearest neighbors in the 2D lattice.

Finally, we note that the perfect in-phase synchronization of vortex precessions in all nanosquares of this 2D system (accompanied by the perfect $~ N^2$ power scaling) illustrates once more the very important point emphasized at the end of Sec. III: there is no phase shift between the oscillations of different STNOs in our system, because the synchronization is achieved due to the magnetodipolar interaction.

\section{VI. CONCLUSION}

We predict, that in various systems consisting of several vortex-based STNOs driven by a spin-polarized current through point contacts attached to square-shaped nanoelements, a very stable synchronization can be achieved. Using a general numerical method to calculate the coupling energy $E_{\rm coup}$ of synchronized STNOs suggested by us, we show that in such a system $E_{\rm coup}$ can be up to three orders of magnitude higher than $kT$, so that the influence of thermal fluctuations is expected to be fully negligible. Further, we have shown that for a system of partially and fully connected nanosquares the strong increase of the coupling constant is not due to the spin-wave exchange between the vortex STNOs, but due to the eliminating of surface 'magnetic charges' of the same sign on adjacent nanosquare edges. This means, that in all kinds of vortex STNOs studied by us, the magnetodipolar interaction plays the dominant role in the establishing of the synchronized regime. 

Finally, we demonstrate, that a robust synchronization of an {\it arbitrary} number of vortex STNOs can be easily obtained employing our concept both in 1D and 2D arrays of such oscillators. This synchronization remains perfectly stable also in a system with a significant distribution of point contact diameters, what is (experimentally unavoidable). Another important advantage of our design is the absence of any phase shift between oscillations of different STNOs (because the synchronization is made possible by the magnetodipolar interaction), so that the perfect $~ N^2$ scaling of the signal power with the number of oscillators $N$ is obtained.

\bibliography{Magnetization_Dynamics}

\end{document}